\documentclass[twocolumn,english,showpacs,showkeys,superscriptaddress,citeautoscript]{revtex4-2}
\usepackage[T1]{fontenc}
\usepackage[latin9]{inputenc}
\usepackage[active]{srcltx}
\usepackage{array}
\usepackage{multirow}
\usepackage{amsmath}
\usepackage{amssymb}
\usepackage{graphicx}
\usepackage{wasysym}
\usepackage{esint}

\makeatletter

\newcommand{\lyxmathsym}[1]{\ifmmode\begingroup\def\b@ld{bold}
  \text{\ifx\math@version\b@ld\bfseries\fi#1}\endgroup\else#1\fi}

\providecommand{\tabularnewline}{\\}

\usepackage{babel}
\usepackage{rotating}

\newcommand{\updwn}{\rotatebox[origin=c]{90}{$\leftrightharpoons$}}
\DeclareMathOperator*{\Motimes}{\text{\raisebox{0.25ex}{\scalebox{0.6}{$\bigotimes$}}}}

\newcommand{\dbar}{{\mathchar'26\mkern-12mu\mathrm{d}}}

\makeatother

\usepackage{babel}
\begin{document}
\title{Quantum Otto engine mimicking Carnot near pseudotransitions in the
one-dimensional extended Hubbard model in the atomic limit}
\author{Onofre Rojas}
\address{Department of Physics, Institute of Natural Science, Federal University
of Lavras, 37200-900 Lavras-MG, Brazil}
\author{Moises Rojas}
\address{Department of Physics, Institute of Natural Science, Federal University
of Lavras, 37200-900 Lavras-MG, Brazil}
\author{S. M. de Souza}
\address{Department of Physics, Institute of Natural Science, Federal University
of Lavras, 37200-900 Lavras-MG, Brazil}
\begin{abstract}
The one-dimensional extended Hubbard model (EHM) in the atomic limit
has recently been found to exhibit a curious thermal pseudo-transition
behavior, which closely resembles first and second-order thermal phase
transitions. This phenomenon, occurring at half-filling, is influenced
by the quantum phase transition between the alternating pair (AP)
and paramagnetic (PM) phases at zero temperature. In this study, we
leverage this anomalous behavior to investigate the performance of
quantum many-body machines, using the EHM as the working substance.
Our analysis reveals that the quantum Otto engine, when operating
in the anomalous region, closely mimics the ideal Carnot engine. In
this region, both the work output and thermal efficiency of the Otto
engine increase, approaching the performance of a Carnot engine. This
highlights the potential of many-body systems, such as the EHM, in
enhancing quantum thermodynamic performance. Our findings demonstrate
that, although the second law of thermodynamics prevents engines from
surpassing Carnot efficiency, the Otto engine can operate remarkably
close to this limit in the anomalous region, offering insights into
new directions for future research on quantum thermodynamic cycles
and working substances.
\end{abstract}
\maketitle

\section{Introduction}

In the past two decades, significant efforts have been made to explore
thermodynamic processes where quantum features of matter are crucial.
This research has successfully combined theoretical proposals \citep{Quan}
with experimental evidence \citep{myers}, bridging the gap between
quantum thermodynamics concepts and practical applications. Quantum
heat engines, like Carnot and Otto, have been studied, and their characteristics
have been compared to classical counterparts. Additionally, a quantum
heat engine model has been analyzed for efficiency and irreversibility
\citep{kasloff,alicki}. Quantum thermodynamics offers a framework
for understanding and optimizing engines and refrigerators \citep{kasloff-14}.
The study of quantum heat engines based on harmonic oscillators has
enhanced our understanding of optimization and irreversibility \citep{insinga,rezek}. 

Recent investigations have advanced quantum Carnot, Otto, and Stirling
engines, surpassing previous limitations in achieving Carnot efficiency
with quantum and nanoscale heat engines. These include achieving Carnot
efficiency through semi-local thermal operations with dual bath interactions
\citep{bera}. Classical magnetic Otto cycles outperformed quantum
ones in work extraction due to thermodynamic equilibrium \citep{pena}.
Energy quantization effects in the Otto cycle were found to enhance
performance, with experimental proposals using trapped ions \citep{Gelbwaser-Klimovsky}.
Several works have explored theoretical dimer systems as working substances,
including quantum engines, refrigerators, and heaters within a quantum
Otto cycle with coupled spins \citep{Oliveira,Makouri,abd}, and including
Dzyaloshinskii--Moriya interactions \citep{asadian}. Though single-spin
and spin lattice systems have been extensively studied \citep{JZHe,Bender},
trimer systems have been rarely explored, with one notable quantum
Otto cycle analysis \citep{aydinger}.

Several quantum machines have been explored using one-dimensional
models. In this context, a quantum Otto heat engine based on a multiferroic
chain as the working substance was investigated in \citep{azimi}.
Halpern et al. \citep{halpern} proposed harnessing many-body localization
for thermodynamic tasks, designing an Otto engine with mesoscale parallel
reliability. In \citep{piccitto}, the thermodynamic cycle of an Otto
engine with a quantum Ising chain as the working substance was examined.
This cycle involves sweeps of the transverse magnetic field in thermal
isolation, alternated with thermalization strokes involving reservoirs
at different temperatures. The performance of quantum thermal machines
using many-body quantum systems or spin-chain models, periodically
connected to external baths via local couplings, was discussed in
\citep{Centamori}. More recently, the spin-chain-star shape model
has been analyzed as the working substance of a quantum Otto cycle
in \citep{Alsulami}. Additionally, researchers have studied a finite-time
quantum Otto cycle in a 1D Bose gas, analyzing efficiency-power trade-offs
\citep{nautiyal}. Nautiyal et al. \citep{arezzo} examined an imperfect
quantum Otto engine, evaluating work, power, and optimal operating
points under less ideal conditions.

The Hubbard model \citep{Hubbard} is fundamental in studying strongly
correlated electrons, particularly where interactions are crucial.
The one-dimensional extended Hubbard model (EHM), which includes nearest-neighbor
interaction energy, is widely studied for its relevance to materials
like carbon nanotubes and organic conductors\citep{Wang-09}. It serves
as a prototype for strongly correlated electron systems \citep{Dagotto}.
Tsuchiizu and Furusaki \citep{Tsuchiizu} revisited its ground-state
phase diagram, revealing new phases and transitions. Glocke et al.
\citep{Glocke} studied thermodynamics, highlighting phase transitions
through compressibility and susceptibility. The phase diagram of the
one-dimensional $U-V$ model at quarter-filling shows a transition
from Luttinger liquid to charge density wave insulator, influenced
by superconducting or spin density wave fluctuations\citep{Zotos}.
In the 1970s, researchers \citep{bari,beni-pincus,gallinar} used
the transfer matrix to study EHM thermodynamics properties such as
specific heat, magnetic susceptibility, and density correlations at
half and quarter fillings with infinite repulsion. Later, Mancini
and Mancini\citep{mancini05,Mancini} identified four phases, charge
orderings, and critical polarization fields \citep{mancini09}.

Quasi-one-dimensional (quasi-1D) materials exhibit fascinating electronic
properties due to reduced dimensionality. For example, $\beta\lyxmathsym{\textquoteright}-\mathrm{EtMe}{{}_3}\mathrm{Sb[Pd(dmit}){{}_2}]{{}_2}$
is a molecular triangular lattice system proposed as a quantum spin
liquid candidate with quasi-1D spin dynamics \citep{Dressel}. Organic
charge-transfer salts like $\mathrm{(TMTTF)}{{}_2}\mathrm{X}$ and
$\mathrm{(TMTSF)}{{}_2}\mathrm{X}$ show strong electron correlations
and charge-ordering phenomena\citep{Riedl}. The TTF finite-energy
spectral features in photoemission of TTF--TCNQ are explained by
a Hubbard-chain description, revealing that $U$ is larger for TTF
than TCNQ\citep{bozi}. One-dimensional electronic transport at the
organic charge-transfer interfaces under high pressure\citep{kang}.
Layered systems like $\mathrm{SrCuO}{{}_2}$ reveal spinon-holon separation,
further emphasizing low-dimensional physics\citep{Oshima}. These
materials collectively highlight dimensionality's role in emergent
quantum phenomena. Additionally, angle-resolved photoemission spectroscopy
(ARPES) studies on 1D extended Hubbard model systems have provided
valuable insights into electron-phonon coupling effects, further validating
theoretical predictions\citep{Hx-Wang}.

The present work is organized as follows: In Sec. 2, we summarize
the thermodynamics of the one-dimensional EHM in the atomic limit,
particularly around anomalous region. In Sec. 3, we discuss thermodynamic
process of quantum many body machines. In Sec. 4, we discuss our results
regarding quantum machine operation modes influenced by pseudo-transitions.
Finally, in Sec. 5, we present our conclusions and perspectives.

\section{1D EHM in the atomic Limit and its thermodynamics}

Therefore, we summarize the most relevant results obtained in reference
\citep{Hubb-ch24}. Despite the simplest version of the EHM has been
extensively studied across various physical systems, our focus is
on a one-dimensional EHM in the atomic limit, where the hopping term
is neglected, resulting in the following Hamiltonian
\begin{alignat}{1}
\boldsymbol{H}= & \sum_{i=1}^{N}\left[U\boldsymbol{n}_{i,\uparrow}\boldsymbol{n}_{i,\downarrow}+V\boldsymbol{n}_{i}\boldsymbol{n}_{i+1}\right.\nonumber \\
 & \left.-\mu(\boldsymbol{n}_{i,\uparrow}+\boldsymbol{n}_{i,\downarrow})-h(\boldsymbol{n}_{i,\uparrow}-\boldsymbol{n}_{i,\downarrow})\right].
\end{alignat}
Here, $U$ denotes the on-site Coulomb interaction between electrons
occupying the same lattice site, $V$ represents the interaction between
electrons on neighboring sites, $\mu$ is the chemical potential that
governs the electron filling of the lattice, and $\boldsymbol{n}_{i,\sigma}$
represents the number operator at site $i$ with spin $\sigma=\{\uparrow,\downarrow\}$
, where $\boldsymbol{n}_{i}=\boldsymbol{n}_{i,\uparrow}+\boldsymbol{n}_{i,\downarrow}$.
The last term accounts for the contribution of the external magnetic
field $h$. 

\subsection{Transfer matrix results\label{subsec:Transfer-matrix}}

To express the transfer matrix of the model, we use the basis $\{|0\rangle,|\uparrow\rangle,|\downarrow\rangle,|\updwn\rangle\}$.
Here, $|0\rangle$ represents the vacuum state, $|\uparrow\rangle$
denotes spin-up, $|\downarrow\rangle$ denotes spin-down, and $|\updwn\rangle$
corresponds to two opposite spins on the same site.

The model can be solved using the transfer matrix technique\citep{beni-pincus,Hubb-ch24},
with the transfer matrix given by
\begin{equation}
\mathbf{W}=\left[\begin{array}{cccc}
1 & yw_{0,1} & y^{-1}w_{0,1} & w_{0,2}\\
yw_{0,1} & y^{2}w_{1,1} & w_{1,1} & yw_{1,2}\\
y^{-1}w_{0,1} & w_{1,1} & y^{-2}w_{1,1} & y^{-1}w_{1,2}\\
w_{0,2} & yw_{1,2} & y^{-1}w_{1,2} & w_{2,2}
\end{array}\right],\label{eq:V}
\end{equation}
where $w_{0,1}={\rm e}^{\beta\mu/2}$, $w_{0,2}={\rm e}^{\beta(\mu-U/2)}$,
$w_{1,1}={\rm e}^{\beta(\mu-V)}$, $w_{1,2}={\rm e}^{\beta(3\mu/2-2V-U/2)}$,
$w_{2,2}={\rm e}^{\beta(2\mu-4V-U)}$, $y={\rm e}^{\beta h/2}$, and
$\beta=1/(k_{B}T)$.

The eigenvalues of the transfer matrix consist of one trivial null
eigenvalue, while the remaining eigenvalues are determined by the
following cubic secular equation
\begin{equation}
\lambda^{3}+a_{2}\lambda^{2}+a_{1}\lambda+a_{0}=0,\label{eq:cub-eq}
\end{equation}
where the coefficients are given by:
\begin{alignat}{1}
a_{0}= & z\left(w_{1,2}^{2}+w_{0,2}^{2}w_{1,1}+w_{0,1}^{2}w_{2,2}\right)\nonumber \\
 & -z\left(w_{1,1}w_{2,2}-2w_{0,2}w_{0,1}w_{1,2}\right),\nonumber \\
a_{1}= & z\left(w_{1,1}+w_{1,1}w_{2,2}-w_{0,1}^{2}-w_{1,2}^{2}\right)+w_{2,2}-w_{0,2}^{2},\nonumber \\
a_{2}= & -1-zw_{1,1}-w_{2,2}.\label{eq:coefs}
\end{alignat}
Therefore, the roots of the algebraic cubic equation \eqref{eq:cub-eq}
can be conveniently expressed using trigonometric functions as follows
\begin{equation}
\lambda_{j}=2\sqrt{Q}\cos\left(\tfrac{\phi-2\pi j}{3}\right)-\frac{1}{3}a_{2},\quad j=0,1,2,\label{eq:sol-cub}
\end{equation}
where
\begin{alignat}{1}
\phi & =\arccos\left(\tfrac{R}{\sqrt{Q^{3}}}\right),\\
Q & ={\displaystyle \frac{a_{2}^{2}-3a_{1}}{9},}\label{eq:Q real}\\
R & ={\displaystyle \frac{9a_{1}a_{2}-27a_{0}-2a_{2}^{3}}{54}.}\label{eq: R real}
\end{alignat}
Additional details regarding the solution, with the largest eigenvalue
$\lambda_{0}$, can be found in reference \citep{Hubb-ch24}.

\subsection{Thermodynamic quantities \label{subsec:Thermo}}

To analyze thermodynamic quantities, we use the grand partition function
for a chain of $N$ sites: 
\begin{equation}
\Xi(T,\mu,h,N)=\lambda_{0}^{N}+\lambda_{1}^{N}+\lambda_{2}^{N},\label{eq:part-func}
\end{equation}
where the eigenvalues are ordered as $\lambda_{0}>\lambda_{1}>\lambda_{2}$.
The grand potential per site in the thermodynamic limit $N\to\infty$,
is determined by the largest eigenvalue of the transfer matrix:
\begin{alignat}{1}
\Omega(T,\mu,h)= & -k_{B}T\ln\lambda_{0}.\label{grand-pot}
\end{alignat}
We define some thermodynamic quantities such as entropy ${\cal S}=-\frac{\partial\Omega}{\partial T}$,
enthalpy $\mathcal{E}=k_{B}T^{2}\frac{\partial\ln\left(\lambda_{0}\right)}{\partial T}$,
magnetization $M=-\frac{\partial\Omega}{\partial h}$, and electron
density $\rho=-\frac{\partial\Omega}{\partial\mu}$. 

\subsection{Frustrated and alternating pair phase transition}

Here we focus on two ground state phases at half-filling band $\rho=1$.
The first phase is the frustrated phase $FR_{2}$ given by
\begin{alignat}{1}
|FR_{2}\rangle= & \Motimes_{j=1}^{N}|\updownarrow\rangle,\nonumber \\
= & \tfrac{1}{\sqrt{z}^{N}}\Motimes_{j=1}^{N}\left(y|\uparrow\rangle+y^{-1}|\downarrow\rangle\right).
\end{alignat}
Its ground-state energy is
\begin{alignat}{1}
E_{FR_{2}}= & V-h-\mu.
\end{alignat}
 The corresponding residual entropy is $\mathcal{S}=k_{B}\ln(2)$.
This state becomes non-frustrated when $h\ne0$, aligning with the
magnetic field.

The other phase is the alternating pair phase ($AP$) expressed by
\begin{alignat}{1}
|AP\rangle= & \Motimes_{j=1}^{N/2}|0,\updwn\rangle\quad\text{or}\quad\Motimes_{j=1}^{N/2}|\updwn,0\rangle.
\end{alignat}
Its ground-state energy is
\begin{alignat}{1}
E_{AP}= & \frac{U}{2}-\mu.
\end{alignat}
The AP phase has no residual entropy and can be identified as a charge
density wave (CDW)\citep{Glocke}.

The ground-state phase boundary between both phases is given by $h=V-\frac{U}{2}$. 

\subsection{Pseudo-critical temperature condition \label{subsec:Psd-crt}}

Recent studies on one-dimensional models with short-range interactions
have revealed pseudo-transitions \citep{pseudo}, which can be analyzed
through spin correlations \citep{Isaac}. These pseudo-transitions
have been observed in several models, including the Ising diamond\citep{Strecka-ising,W-Yin-2},
ladder \citep{hutak21,chapman,W-Yin-1}, Ising-Heisenberg diamond
\citep{torrico,torrico2}, double-tetrahedral\citep{Galisova}, and
Ising-Heisenberg ladder models \citep{on-strk}.

In the context of ground-state phase transitions, we summarize the
thermal behavior in the low-temperature region, as explored in Ref.
\citep{Hubb-ch24}.

At low temperatures, an anomalous behavior is observed at a characteristic
temperature, referred to as the pseudo-critical temperature $T_{p}$
\citep{Hubb-ch24}, which satisfies the following relation

\begin{equation}
w_{1,1}z-w_{0,2}=0.\label{eq:w-cnd-0}
\end{equation}
Explicitly, with $w_{0,2}={\rm e}^{\beta(\mu-U/2)}$, $w_{1,1}={\rm e}^{\beta(\mu-V)}$,
and $z=2\cosh(\beta h)$, we have the following expression 
\begin{equation}
2\cosh(\beta_{p}h)={\rm e}^{\beta_{p}(V-U/2)}.\label{eq:w-cond-h}
\end{equation}
Particularly, for $0<h\lesssim V-\frac{U}{2}$, an anomalous behavior
arises, indicating a pseudo-transition at temperature $T_{p}$.

Although Eq.\eqref{eq:w-cond-h} is independent of the chemical potential
$\mu$, this condition is valid only within a limited interval, $1<\mu<2$.
For details, refer to the phase diagram in Ref. \citep{Hubb-ch24}. 

The sharp, yet continuous, entropy jump near the pseudo-critical temperature
$T_{p}$, although not signaling a true phase transition, reveals
a rapid microscopic reconfiguration in the system. This enhanced response
suggests that even minor temperature variations can lead to substantial
changes in the working substance's thermal properties. Notably, $T_{p}$
is unrelated to a true critical temperature, yet studies \citep{hutak21,unv-cr-exp,krokhmalskii21}
suggest universality in pseudo-critical exponents, adhering to the
Rushbrooke inequality. Similar behavior appears in the extended Hubbard,
Potts \citep{psd-hub-dmd}, Zimm-Bragg-Potts \citep{Potts-psd}, and
spin-pseudospin models \citep{Yasinkskaya,karlova}.

\subsection{Magnetic Properties}

Here, we examine the magnetic influence near the pseudo-transition.
At low temperatures, quantum phase transitions often enhance the magnetocaloric
effect, a well-studied phenomenon. Following this perspective, we
analyze low-temperature isentropic curves in regions near quantum
phase transitions occur at zero temperature.

\begin{figure}
\includegraphics[scale=0.56]{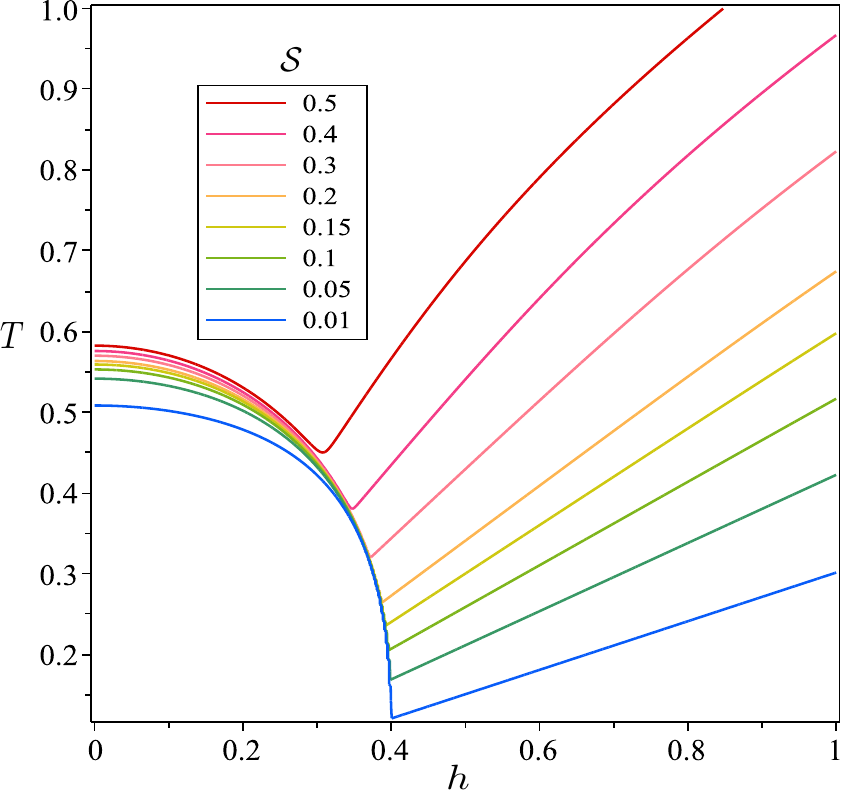}

\caption{\label{fig:Isentropic}Isentropic curve around anomalous region, in
the plane $h-T$, assuming $U=10$, $V=5.4$, $\mu=15.8$}
\end{figure}

\begin{figure}
\includegraphics[scale=0.64]{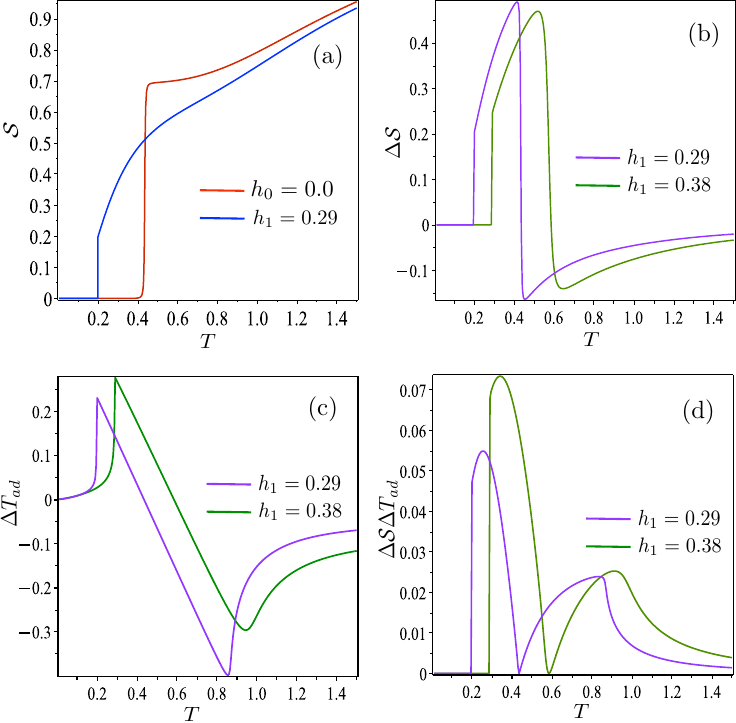}\caption{\label{fig:Entrps}(a) Entropy as a function of temperature $T$,
with parameters $U=10$, $V=5.3$ and $\mu=15.8$. (b) Entropy variation
$\Delta\mathcal{S}=\mathcal{S}(T,h_{1})-\mathcal{S}(T,0)$ as a function
of temperature $T$, with fixed parameters $U=10$, $\mu=15.8$, and
$V=5.3$, for $h_{1}=0.29$ (violet curve). Similarly, for $h_{1}=0.38$
(green curve) and fixed parameters $U=10$, $\mu=15.8$, and $V=5.4$.
(c) Adiabatic temperature change $\Delta T_{ad}$ as a function of
temperature, for the same set of parameters as in (b). (d) The product
$\Delta\mathcal{S}\Delta T_{ad}$ as a function of temperature, under
the same conditions as in (b).}
\end{figure}

In Fig. \ref{fig:Isentropic}, the isentropic curve is illustrated
in the $h-T$ plane, considering the following fixed parameters: $U=10$,
$V=5.4$, and $\mu=15.8$. The isentropic curve exhibits unusual behavior
over a certain range of entropy, as shown in the legend. Specifically,
for $0.3\apprle h<0.4$, all isentropic curves are nearly overlapping.
This unusual behavior significantly affects the thermal properties
and, for lower magnetic fields, the curves remain densely packed,
highlighting the influence of the anomalous region. The blue curves
delineate the quasi-alternating pair ($qAP$) region for $h\lesssim0.4$,
whereas for $h\gtrsim0.4$, the area below the blue curve corresponds
to the quasi-paramagnetic ($qPM$) region. Additionally, for a null
magnetic field and at temperatures roughly around $T\sim0.5-0.6$,
there is evidence of a strong change in entropy, indicating a transition
from the $qAP$ to the $qFR_{2}$ region. In Fig. \ref{fig:Isentropic},
below blue curve ($T\lesssim0.01$) basically the entropy is null,
see Fig. \ref{fig:Entrps}a from another perspective.

Fig. \ref{fig:Isentropic} shows unusual thermodynamic behavior, with
nearly overlapping isentropic curves indicating an enhanced magnetocaloric
effect in the pseudo-critical region. Despite the absence of singularities,
sharp entropy redistribution occurs, potentially benefiting quantum
thermodynamic cycles by enabling efficient thermal control without
fine-tuned critical conditions.

For this purpose, in fig.\ref{fig:Entrps}a, the entropy is shown
as a function of temperature, assuming $U=10$, $V=5.3$ and $\mu=15.8.$
For a zero magnetic field and $h_{1}=0.29$, we observe a typical
abrupt yet continuous change in entropy. This jump occurs for $h_{0}=0$
at $T\sim0.2$ and for $h_{1}=0.29$, the jump occurs at $T\sim0.43$.
This continuous jump mimics a first order phase transition; however,
it is important to note that no true phase transition takes place.

In panel (b), we illustrate the variation of entropy $\Delta\mathcal{S}=\mathcal{S}(T,h_{1})-\mathcal{S}(T,0)$
as a function of temperature. We assume $U=10$, $\mu=15.8$ and $V=5.3$,
with $h_{1}=0.29$ and $h_{0}=0$. The entropy variation $\Delta\mathcal{S}$
is represented by the violet curve. Below the jump, $\Delta\mathcal{S}$
is very small, and a strong jump occurs at $T\sim0.2$, leading to
a large $\Delta\mathcal{S}$, followed by a strong change that results
in a negative value at $T\sim0.43$, after which it increases gradually.
A similar behavior is observed for $U=10$, $\mu=15.8$, and $V=5.4$,
with $h_{1}=0.38$ and $h_{0}=0$, where $\Delta\mathcal{S}$ is represented
by the green curve. The jump in $\Delta\mathcal{S}$ is again small,
with a strong jump occurring at $T\sim0.29$, leading to a large $\Delta\mathcal{S}$,
followed by a strong change that results in a negative value at $T\sim0.56$,
and then increasing gradually for higher temperatures.

In panel (c), the adiabatic temperature change for the same set of
parameters as in panel (b) is shown, which is given by 
\[
\Delta T_{ad}=-\int_{h_{0}}^{h_{1}}\left(\frac{\partial T}{\partial\mathcal{S}}\right)_{h}\left(\frac{\partial M}{\partial T}\right)_{h}{\rm d}h,
\]
 as a function of temperature. In practice, we can determine $\Delta T_{ad}$
by solving the following equation $\mathcal{S}(T+\Delta T_{ad},h_{1})=\mathcal{S}(T,h_{0})$.
To compare our results, we use the same curve colors as in panel (b).
The violet curve exhibits a sharp peak at $T\sim0.2$, which decreases
almost linearly, reaching zero around $T\sim0.43$, followed by a
minimum at approximately $T\sim0.85$, after which increases monotonically.
A similar behavior is observed for the green curve, with a sharp peak
at $T\sim0.29$, a minimum at $T\sim0.95$, and a monotonic increase
for higher temperatures.

Panel (d) presents a further analysis of the quantity $\Delta\mathcal{S}\Delta T_{ad}$,
which could be particularly relevant for the operation of a Carnot
engine. In this analysis, we continue using the same set of parameters
established in panel (b). Notably, for temperatures below $T\approx0.2$
(violet) and $T\approx0.29$ (green), $\Delta\mathcal{S}\Delta T_{ad}\sim0$.
Then, a significant amount of $\Delta\mathcal{S}\Delta T_{ad}$ appears,
vanishing at $T\sim0.44$ and $T\sim0.58$, respectively. For higher
temperatures, the quantity increases, reaching a maximum at $T\sim0.84$
and $T\sim0.91$, respectively. After these peaks, the quantity begins
to gradually decrease.

The entropy variation $\Delta\mathcal{S}$ in Fig. \ref{fig:Entrps}b
reveals a sharp yet continuous crossover, mimicking a first-order
transition without true discontinuities. This pseudo-transition reflects
a rapid microscopic reconfiguration without singularities in response
functions, ensuring stable behavior across all temperatures. Such
controlled entropy changes could be useful for quantum thermodynamic
processes.

\section{Reversible cycle Process}

In the following, we proceed to analyze the thermal machine cycles.
Particularly, we will focus on the reversible Carnot and Otto engines.

\subsection{Work and heat transferred}

The grand potential, denoted as $\Omega$ and defined by \eqref{grand-pot},
is a key quantity for analyzing systems in the grand canonical ensemble.
This ensemble describes systems where the particle number can fluctuate,
and external magnetic field act as control parameters. The first law
of thermodynamics for the present system can be expressed as:
\begin{equation}
{\rm d}{\cal E}=T{\rm d}{\cal S}-M{\rm d}h-\rho{\rm d}\mu,\label{eq:d-entlp-1}
\end{equation}
where $\dbar Q=T{\rm d}{\cal S}$ is the heat absorbed by the system,
and $\dbar W=M{\rm d}h+\rho{\rm d}\mu$ is the work done on the surroundings
by the system.

Here, the quantities ${\cal E}$ (enthalpy), ${\cal S}$ (entropy),
$M$ (magnetization), and $\rho$ (particle density) are defined per
lattice site to normalize extensive variables in systems with fixed
lattice size.

\subsection{Operation modes of machine}

Considering the one-dimensional EHM in the atomic limit as the working
substance, the work is linked to changes in energy levels as the magnetic
field $h$ is altered. By adjusting the parameters of the reversible
cycle in a 1D EHM in the atomic limit, the total work of the cycle
can be reversed in accordance with the second law of thermodynamics,
which, as formulated by Clausius, defines operational regions for
thermodynamically reversible cycles based on temperature differences
($T_{h}>T_{l}$). In heat engine mode, the system absorbs heat from
a hotter bath at $T_{h}$ ($Q_{in}>0$) and releases it to a cooler
bath at $T_{l}$ ($Q_{out}<0$), partially converting the heat into
work ($W_{net}>0$), as shown in Table \ref{tab:engines}. In refrigerator
mode, the model induces heat flow from the cooler bath at $T_{l}$
to the hotter bath at $T_{h}$, requiring more work input than work
output, resulting in net work performed ($W_{net}<0$). As a thermal
accelerator, the system uses net work ($W_{net}<0$) to increase heat
flow from the hotter to the cooler bath. In heater mode, the net work
($W_{net}<0$) is applied to generate heat flow in both baths, causing
the 1D EHM in the atomic limit to release heat in both baths ($Q_{in}<0$
and $Q_{out}<0$). All these processes are summarized in Table \ref{tab:engines}.
In particular, we apply the Carnot and Otto engine processes as detailed
in Appendix \ref{sec:Engine-Process}.

\begin{table}
\caption{\label{tab:engines}Characteristics of work and heat as allowed by
the second law of thermodynamics. The sign ($+$) indicates work done
by the system and heat absorbed, while the sign ($-$) indicates work
done on the system and heat released.}
\begin{tabular}{|l|c|c|c|c|}
\hline 
Operation mode & $W_{net}$ & $Q_{in}$ & $Q_{out}$ & \multirow{1}{*}{Thermal efficiency}\tabularnewline
\hline 
\hline 
Heat engine & $+$ & $+$ & $-$ & $\eta=\frac{W_{net}}{Q_{in}}$\tabularnewline
\hline 
Refrigerator & $-$ & $-$ & $+$ & $COP=\frac{Q_{in}}{W_{net}}$\tabularnewline
\hline 
Heater & $-$ & $-$ & $-$ & $COP=\frac{Q_{out}}{W_{net}}$\tabularnewline
\hline 
Accelerator & $-$ & $+$ & $-$ & $COP=\frac{Q_{out}}{W_{net}}$\tabularnewline
\hline 
\end{tabular}
\end{table}

Fig. \ref{fig:Entrps}c shows the adiabatic temperature change $\Delta T_{ad}$,
essential for the Otto cycle. The sharp entropy variation in the pseudo-transition
region enhances the thermodynamic response, where small parameter
changes significantly impact work and efficiency. This suggests that
quantum engines could leverage this region for optimal performance
without critical divergences. The smooth yet sharp entropy change
enables near-reversible cycles, allowing the Otto cycle to closely
mimic Carnot efficiency.

\subsection{Thermal efficiency}

In the heat engine mode, the efficiency of the quantum engine using
the 1D EHM in the atomic limit as the working substance is presented
in Table \ref{tab:engines}, where $\eta=\frac{W_{net}}{Q_{in}}<1$.
For the other operational modes, the coefficient of performance (COP)
is used, as detailed in Table \ref{tab:engines}. Since COP values
are typically greater than one, we introduce an alternative metric,
\begin{equation}
\kappa=\frac{COP}{1+COP},\label{eq:kp}
\end{equation}
 which maps the COP to a restricted range of $0<\kappa<1$. This metric
implies that as COP approaches $0$, $\kappa$ also approaches $0$,
and as COP tends toward infinity, $\kappa$ approaches $1$. Notably,
a COP of $1$ corresponds to $\kappa=0.5$. As a result, the thermal
efficiency for all operational modes: refrigerator, heater, and accelerator
can be consistently defined within this range, where $0$ indicates
the worst performance and 1 represents the best, as described in Table
\ref{tab:engines}.

For the four-stroke process, the net work performed is given by $W_{net}=W_{ab}+W_{cd}=Q_{bc}+Q_{da}$.
Therefore, the efficiency of the Otto engine is expressed as
\begin{alignat}{1}
\eta_{o}= & \frac{W_{net}}{Q_{bc}}=1+\frac{Q_{da}}{Q_{bc}},
\end{alignat}
since the cyclic process occurs around two pseudo-critical temperatures,
we express it algebraically in this region to illustrate the efficiency
of the Otto engine, as detailed in Appendix \ref{sec:Otto-Machine-efficiency}.

From Appendix \ref{sec:Otto-Machine-efficiency}, the efficiency of
the Otto engine around the pseudo-transition is approximately given
by 
\begin{equation}
\eta_{o}=\eta_{c}-\frac{T_{p_{1}}}{T_{p_{2}}}\left(\frac{\mathcal{S}_{0}+\mathcal{S}_{1}}{2C_{p_{2}}}-\frac{\mathcal{S}_{p_{2}}}{C_{p_{2}}}\right).\label{eq:Otto-effcy}
\end{equation}
As expected, the Otto efficiency is never greater than the Carnot
efficiency, $\eta_{c}\geqslant\eta_{o}$, since the last term is a
positive quantity. However, this term becomes sufficiently small,
leading to $\eta_{o}\rightarrow\eta_{c}$. 

This is an intriguing result, as it demonstrates that by simply following
a four-stroke reversible Otto engine process, one can achieve a near-maximum
efficiency comparable to the Carnot efficiency, without requiring
any additional modifications to enhance performance.

\section{Results of Carnot and Otto machine}

The classical Carnot Cycle, regarded as the most efficient thermodynamic
cycle, contrasts with the Otto Cycle, which idealizes spark-ignition
engines. Recently, Kroetz\citep{Kroetz} reexamined this comparison
didactically, emphasizing the impracticality of the Carnot cycle in
real engines due to its dependency on the compression ratio.

The quantum Carnot and Otto cycles, composed of a four-stroke process,
are explored under general conditions detailed in Appendix \ref{sec:Engine-Process},
where express the explicit expressions for heat absorbed, heat released
and work performed for both engines.

\begin{figure}
\includegraphics[scale=0.7]{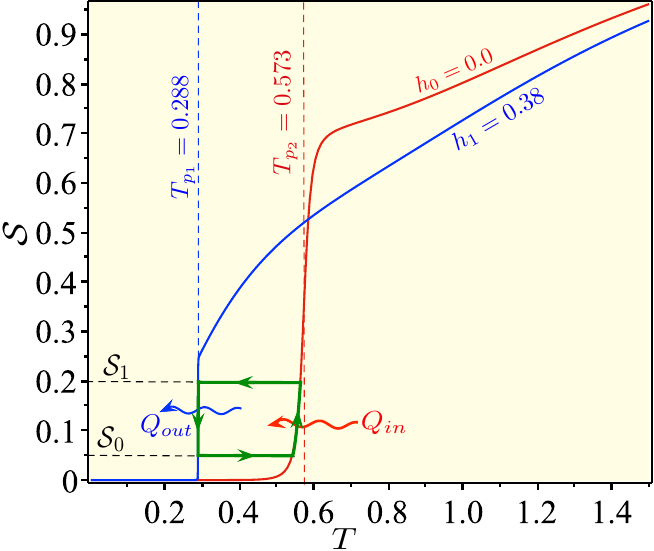}

\caption{\label{fig:Otto-cycle-Hmod}The Otto cycle, which mimics the Carnot
cycle, is executed by the 1D EHM in the atomic limit using the parameters
listed in Table \ref{tab:Ott-parm}, where $\mathcal{S}_{0}=0.05$
and $\mathcal{S}_{1}=0.2$. }

\end{figure}

In Fig. \ref{fig:Otto-cycle-Hmod}, the Otto cycle engine is illustrated
around the anomalous region \citep{Hubb-ch24}. Here, we assume fixed
parameters $\mu=15.8$, $U=10$ and $V=5.4$, satisfying the conditions
of the anomalous region. In this region, we notice an intriguing behavior
of the Otto cycle that mimics the Carnot cycle: the isomagnetic process
behaves almost like an isothermal process (vertical line). By \textquotedblleft isomagnetic''
process we mean a process occurring at a constant magnetic field.
The Otto processes, as described in Appendix \ref{sec:Engine-Process},
can be performed, for instance, by fixing the following two adiabatic
processes at $\mathcal{S}_{0}=0.05$ and $\mathcal{S}_{1}=0.2$, and
two isomagnetic processes at $h_{0}=0$ and $h_{1}=0.38$. In Table
\ref{tab:Ott-parm}, the cold temperature ($T_{l}$), hot temperature
($T_{h}$), and two temperatures corresponding to the pseudo-critical
points ($T_{p_{1}}$ and $T_{p_{2}}$) are reported. Notably, the
specific heat at $T_{p_{1}}$ and $T_{p_{2}}$ is considerably larger
compared to the specific heat at the cold and hot temperatures.

During the isomagnetic process, the system resembles an isothermal
process (represented by an almost vertical line). Consequently, this
cyclic process (green cycle) approaches the behavior of a Carnot engine,
becoming nearly rectangular in shape. The algebraic expression for
the efficiency approximation of this process is detailed in Appendix
\ref{sec:Otto-Machine-efficiency} as an illustrative example. However,
all figures were generated from the exact solution without any approximations.

\begin{table}
\caption{\label{tab:Ott-parm}Illustrative temperatures, entropies and specific
heat, for the 1D EHM in the atomic limit, assuming the parameters:
$\mu=15.8$ and $U=10$.}
\begin{tabular}{|c|c|l|l|l|r|}
\hline 
 & Temperature & $h$ & $V$ & Entropy & Specific heat\tabularnewline
\hline 
\hline 
$T_{l}$ & $0.2883031789$ & $0.38$ & $5.4$ & $0.05$ & $64.5881093600$\tabularnewline
\hline 
$T_{p_{1}}$ & $0.2885145480$ & $0.38$ & $5.4$ & $0.1231215757$ & $123.9985479937$\tabularnewline
\hline 
$T'_{l}$ & $0.2887433779$ & $0.38$ & $5.4$ & $0.2$ & $59.3619717363$\tabularnewline
\hline 
$T'_{h}$ & $0.5416370179$ & $0$ & $5.4$ & $0.05$ & $3.6339831194$\tabularnewline
\hline 
$T_{p_{2}}$ & $0.5730760977$ & $0$ & $5.4$ & $0.3522945133$ & $9.9532683846$\tabularnewline
\hline 
$T_{h}$ & $0.5634744399$ & $0$ & $5.4$ & $0.2$ & $7.3916139289$\tabularnewline
\hline 
\hline 
$T_{l}$ & $0.1964545082$ & $0.29$ & $5.3$ & $0.05$ & $2888.0095116037$\tabularnewline
\hline 
$T_{p_{1}}$ & $0.1964570150$ & $0.29$ & $5.3$ & $0.0987144858$ & $4389.4535640584$\tabularnewline
\hline 
$T'_{l}$ & $0.1976924508$ & $0.29$ & $5.3$ & $0.2$ & $0.4129700921$\tabularnewline
\hline 
$T'_{h}$ & $0.4260966401$ & $0$ & $5.3$ & $0.05$ & $5.3306912666$\tabularnewline
\hline 
$T_{p_{2}}$ & $0.4325649629$ & $0$ & $5.3$ & $0.2722766206$ & $38.0622167546$\tabularnewline
\hline 
$T_{h}$ & $0.4307158885$ & $0$ & $5.3$ & $0.2$ & $28.2638604911$\tabularnewline
\hline 
\end{tabular}
\end{table}
 
\begin{figure}
\includegraphics[scale=0.6]{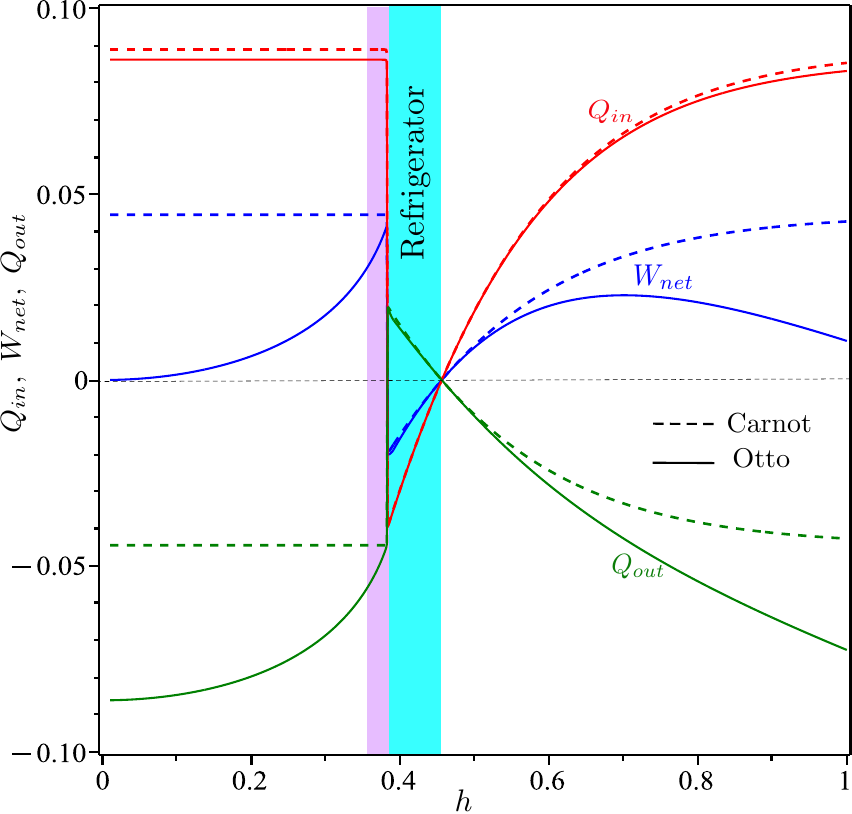}\caption{\label{fig:QQW1}Heat absorbed ($Q_{in}$), released heat ($Q_{out}$),
and work performed ($W_{net}$) as a function of magnetic field, assuming
$U=10$, $V=5.4$, and $\mu=15.8$. Particularly here we fixed the
$T_{h}=0.56$ and $T_{l}=0.28$. }
\end{figure}

To continue analyzing our results, Fig. \ref{fig:QQW1} shows the
heat absorbed (red curve), heat rejected (green curve), and the work
performed (blue curve). The solid line represents the Otto machine
process, while the dashed line corresponds to the Carnot machine.
In plotting the curves, we fixed the cold temperature $T_{l}=0.28$
and the hot temperature $T_{h}=0.56$, varying the external magnetic
field. We observe three distinct regions: Below $h=0.38$, the work
performed is positive, heat absorbed is positive, and heat rejected
is negative. In this region, the working substance (1D EHM in the
atomic limit) operates as a thermal engine. For the Carnot engine,
$Q_{in}$, $Q_{out}$ and $W_{net}$ are independent of $h$ (see
Appendix \ref{sec:Engine-Process}), while for the Otto engine, $Q_{in}$
is almost independent of $h$ (but lower than the Carnot heat absorbed).
Meanwhile, $Q_{out}$ increases with the magnetic field, approaching
the Carnot heat when $h\to0.386$. Similarly, the work performed increases,
reaching the Carnot work performed at $h\to0.386$. In contrast, for
the magnetic field range $0.386<h<0.46$, the system operates as a
refrigerator, as both the work performed ($W_{net}$) and heat absorbed
($Q_{in}$) are negative, while $Q_{out}$ is positive. We also observe
that both the Carnot and Otto refrigerators behave almost identically,
as there is no significant distinction between the solid and dashed
lines. A similar analysis applies for $h>0.46$. In this region, we
have $Q_{in}>0$, $Q_{out}<0$, and $W_{net}>0$, which corresponds
to a thermal engine for both the Carnot and Otto machines.

\begin{figure}
\includegraphics[scale=0.6]{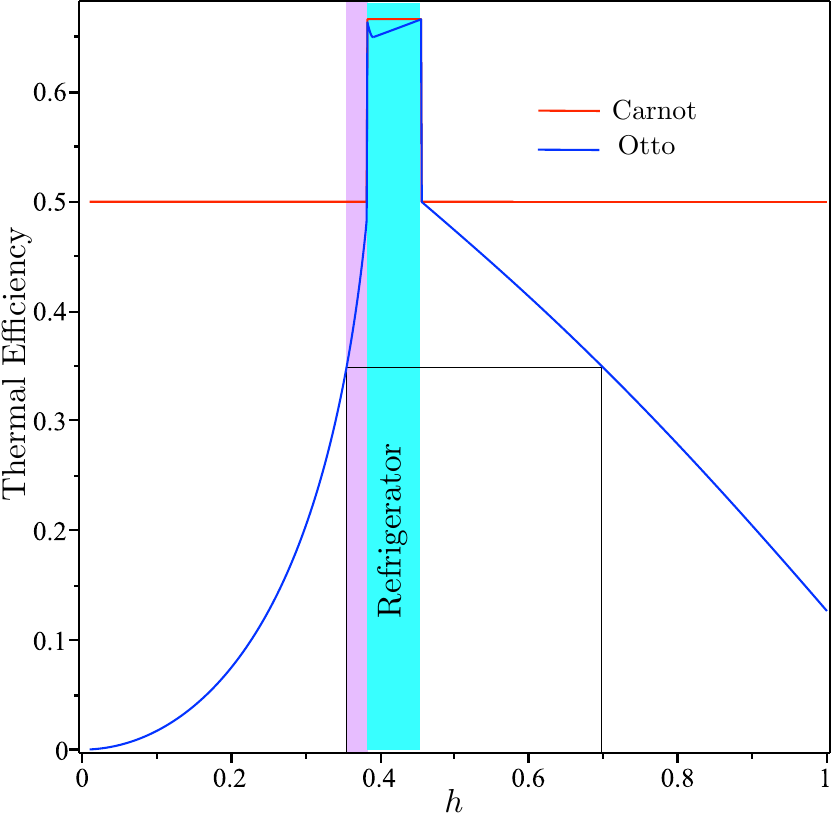}\caption{\label{fig:Effcy}Thermal efficiency of Carnot and Otto engine as
a function of magnetic field, considering $U=10$, $V=5.4$, and $\mu=15.8$.
Particularly here we fixed the $T_{h}=0.56$ and $T_{l}=0.28$. }

\end{figure}

In Fig. \ref{fig:Effcy}, we illustrate the thermal efficiency under
the same conditions as in Fig. \ref{fig:QQW1} to compare the thermal
efficiencies of the Carnot and Otto machines, represented by the red
and blue curves, respectively. For $h>0.46$, there is a maximum work
output, which occurs roughly at $h=0.70$, as shown in Fig. \ref{fig:QQW1}.
This point is marked in Fig. \ref{fig:Effcy} by a small line, and
the corresponding horizontal line (representing the same efficiency)
intersects the first region of the thermal engine at approximately
$h\approx0.36$. Therefore, the left side of the violet region corresponds
to a range where the efficiencies of both thermal engines are equivalent.
In this violet region, the work performed by the Otto engine and its
thermal efficiency increase, approaching the work and efficiency of
the Carnot engine.

This is an interesting result since our working substance (1D EHM
in the atomic limit) becomes highly efficient in the Otto engine ($\eta_{o}\rightarrow0.5$),
nearly matching the Carnot engine efficiency ($\eta_{c}=0.5$). On
the other hand, when the system operates as a refrigerator, both the
Carnot and Otto cycles exhibit nearly the same thermal efficiency,
as given by \eqref{eq:kp}. However, for higher magnetic fields, the
Otto efficiency decreases.

\begin{figure}
\includegraphics[scale=0.6]{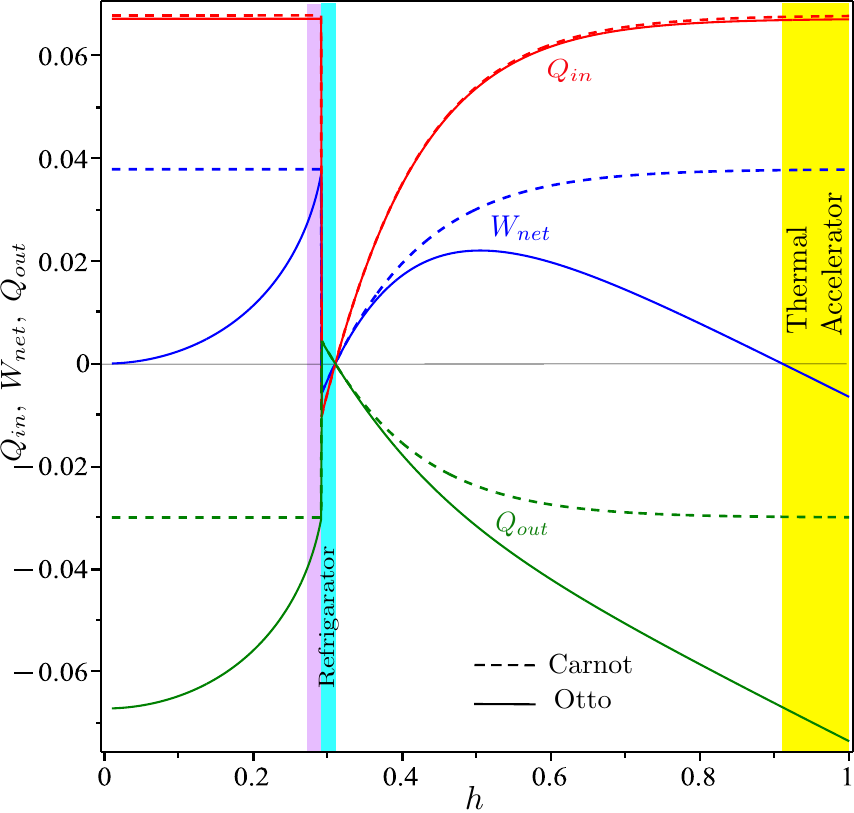}\caption{\label{fig:QQW2}Heat absorbed ($Q_{in}$), heat released ($Q_{out}$),
and work performed ($W_{net}$) as functions of the magnetic field,
assuming $U=10$, $V=5.3$, and $\mu=15.8$. Specifically, here we
have fixed $T_{h}=0.43$ and $T_{l}=0.19$.}
\end{figure}

\begin{figure}
\includegraphics[scale=0.6]{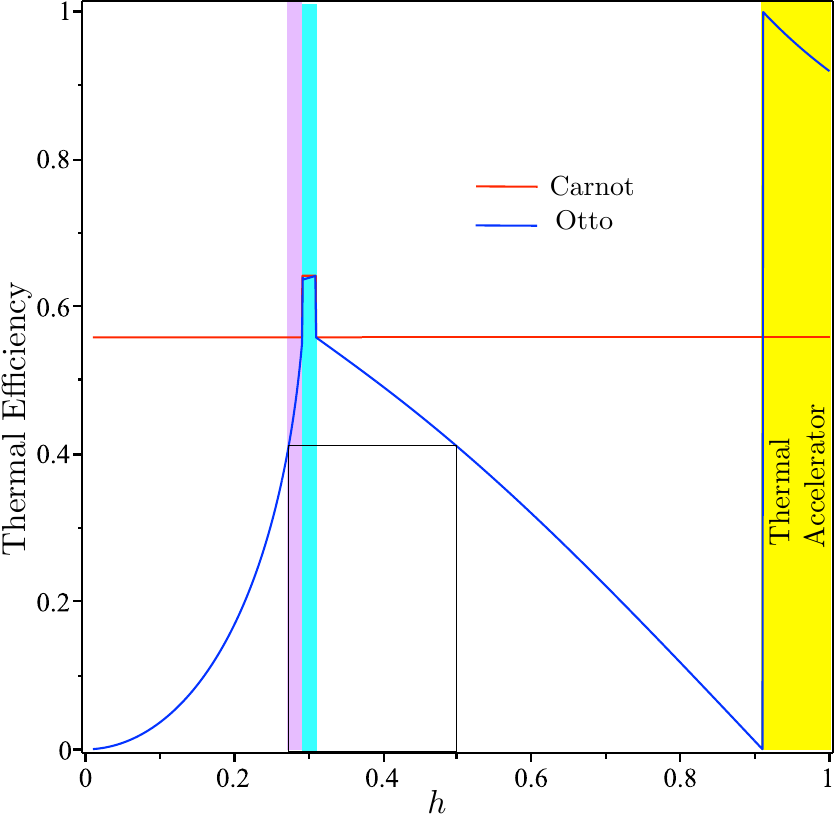} \caption{\label{fig:Effcy2} Thermal efficiency of Carnot and Otto engine as
a function of magnetic field, considering $U=10$, $V=5.3$, and $\mu=15.8$.
Particularly here we fixed the $T_{h}=0.43$ and $T_{l}=0.19$.}
\end{figure}

To corroborate our results, we choose another set of parameters: $U=10$,
$\mu=1.58$ and $V=5.3$, with cold and hot temperatures given by
$T_{l}=0.19$ and $T_{h}=0.43$. Fig. \ref{fig:QQW2} illustrates
a plot similar to Fig. \ref{fig:QQW1} for $Q_{in}$, $Q_{out}$ and
$W_{net}$ as a function of the magnetic field. Once again, we observe
two engine regions: for $0<h<0.29$ and for $0.31<h<0.91$. There
is a region where the system behaves as a refrigerator ($0.29<h<0.31$),
and for higher magnetic fields ($h>0.91$), the system operates as
a thermal accelerator (yellow region). The violet region represents
an area with high performance and a large amount of work performed,
as shown in Fig. \ref{fig:Effcy2}. This region corresponds to the
optimal operating range where the Otto engine achieves its best performance.

Similar to Fig. \ref{fig:Effcy}, Fig. \ref{fig:Effcy2} shows the
thermal efficiency of both the Carnot (red) and Otto (blue) engines
as a function of the magnetic field $h$ (see Eq. \ref{eq:Otto-effcy}),
assuming the same set of parameters as in Fig. \ref{fig:QQW2}. The
Carnot engine efficiency is given by $\eta_{c}=1-T_{l}/T_{h}=0.56$,
while the Otto engine efficiency $\eta_{o}$ depends on the magnetic
field $h$. The cyan region ($0.29<h<0.31$) corresponds to the thermal
efficiency of a refrigerator, while the yellow region ($h>0.91$)
represents the thermal efficiency of a thermal accelerator. The efficiency
$\kappa$ for both is given by Eq. \eqref{eq:kp} and Table \ref{tab:engines}.
For $h=0.5$, the work performed at higher magnetic fields reaches
its maximum in Fig. \ref{fig:QQW2}. This region, marked as the violet
region ($0.26<h<0.29$) in Fig. \ref{fig:Effcy2}, corresponds to
the highest efficiency. Therefore, this violet region corresponds
to a highly efficient regime (very close to Carnot efficiency) with
a significant amount of work performed, indicating that the Otto engine
mimics the Carnot engine.

The near-isothermal behavior observed during the isomagnetic process
where the Otto cycle closely mimics a Carnot engine underscores the
potential for optimized thermal control. Such behavior, characterized
by a pronounced magnetocaloric effect, may provide a strategic advantage
in designing quantum engines with efficiency approaching the Carnot
limit.

Of course, any engine can be manipulated in accordance with the fundamental
Carnot theorem of the second law of thermodynamics to achieve Carnot
efficiency. However, our main result demonstrates a natural way in
which the Otto engine closely mimics the Carnot engine near the anomalous
region. This implies that the Otto engine could approach the performance
of an almost perfect machine.

\section{Conclusion }

This study investigates the theoretical application of the one-dimensional
EHM in the atomic limit, a fundamental physical property initially
studied in the 1970s \citep{bari,beni-pincus,gallinar}. Subsequent
studies, particularly by Glocke \citep{Glocke} and Mancini \citep{Mancini},
revealed its anomalous behavior, which has recently been reexamined
to uncover finite-temperature pseudo-transitions in certain 1D models
\citep{Hubb-ch24}. These pseudo-transitions occur in specific regimes,
offering intriguing insights into quantum many-body systems.

Given the growing interest in quantum many-body machines and their
technological potential, we explored the Carnot and Otto machines
with the 1D EHM in the atomic limit as the working substance. Our
primary focus was the anomalous region, where pseudo-transitions occur,
particularly the transition between the alternating pair (AP) and
paramagnetic (PM) phases \citep{Hubb-ch24}. Near the pseudo-critical
region, our analysis shows that the Otto engine closely mimics the
Carnot engine, with its performance approaching that of an ideal machine.
While both engines operate through a four-stroke cycle, there are
notable distinctions: the Carnot cycle involves two adiabatic and
two isothermal processes performed quasi-statically to maintain fixed
temperatures, whereas the Otto engine uses two isomagnetic processes
(fixed magnetic field) where temperatures are allowed to vary. This
adaptability gives the Otto engine an edge in practical scenarios,
as it combines high efficiency with operational simplicity.

Although the second law of thermodynamics prohibits any engine from
exceeding Carnot efficiency, our findings demonstrate that, within
the anomalous region, the Otto machine mimics quite well Carnot machine.
Note that this process is not occurring in phase transition region
but in an anomalous region only. This underscores the potential of
many-body systems in enhancing quantum thermodynamic performance. 

It is well known that engines operating between the same hot and cold
reservoirs can achieve similar efficiencies, but their practical effectiveness
often depends on physical constraints. For instance, Carnot cycles,
while theoretically ideal, require long time intervals due to the
quasi-static nature of their adiabatic processes. Such long durations
can limit their power output. In contrast, the Otto cycle offers a
practical advantage by enabling faster cycles, making it more efficient
for generating work within shorter timeframes. This highlights the
importance of analyzing and comparing different engine cycles, as
we have done here, to identify trade-offs between efficiency and power
output, similarly that was explored in finite-time thermodynamics\citep{Insinga-mdpi}.

Figs. 1 and 2 illustrate that the pseudo-transition region offers
a unique thermodynamic regime characterized smooth entropy variation,
enabling efficient quantum thermodynamic cycles. Unlike true critical
points, which require fine-tuning or result in divergences, the pseudo-transition
region provides stable thermodynamic control. These findings suggest
that engineered quantum materials exhibiting pseudo-transitions could
be promising candidates for quantum heat engines and other thermodynamic
applications.

While our study focuses on the 1D EHM in the atomic limit, the principles
and findings can likely extend to other systems exhibiting similar
anomalous properties. This paves the way for exploring alternative
theoretical engines and identifying unique advantages offered by different
working substances. The anomalous behavior observed in this model
is just one application, and future work could explore other characteristics
of such systems, potentially revealing new and unexpected behaviors
that challenge conventional thermodynamic paradigms.

Although our analysis is confined to a one-dimensional framework,
the insights gained offer a conceptual stepping stone for extending
these ideas to higher dimensions. In particular, the pronounced response
functions observed here could qualitatively guide experimental investigations
in quasi-1D or even 2D/3D systems, where similar microscopic reconfigurations
might occur under finite-size effects.
\begin{acknowledgments}
O.R., M. R. and S. M. de S. thank Brazilian agencies FAPEMIG and CNPq
for partial financial support. 
\end{acknowledgments}

\appendix

\section{\label{sec:Engine-Process}Carnot and Otto Process}

This appendix provides a comprehensive overview of both Carnot and
Otto engines, detailing their principles, operation, and key characteristics.

\subsection{Quantum many-body Carnot machine cycle}

\begin{figure}
\includegraphics[scale=0.9]{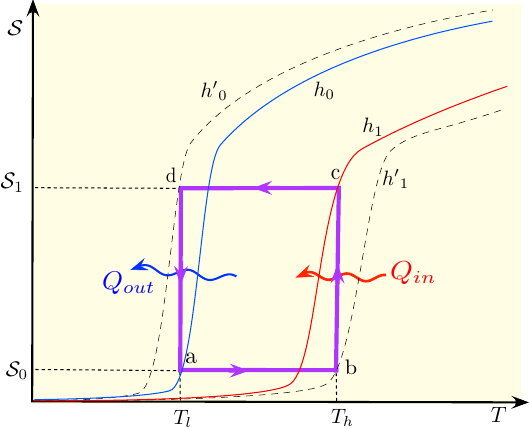}\caption{\label{fig:Carnot-cycl} A schematic representation of a quantum many-body
Carnot machine, assuming $T_{l}=T_{p_{1}}$ and $T_{h}=T_{p_{2}}$.}
\end{figure}

Here, we explore the typical operation of a reversible Carnot engine,
as illustrated in Fig. \ref{fig:Carnot-cycl}.

\begin{enumerate}
\item The process $a\rightarrow b$ is adiabatic, meaning no heat is exchanged
($Q_{ab}=0$). According to a fundamental thermodynamic result, the
work done during this process is given by:
\begin{alignat}{1}
W_{ab}= & -\Delta{\cal E}_{ab}=-({\cal E}_{b}-{\cal E}_{a})\nonumber \\
= & {\cal E}(T{}_{l},h_{0})-{\cal E}(T_{h},h'_{1}),
\end{alignat}
where $h_{0}$ can be found from the adiabatic condition $\mathcal{S}(T_{l},h'_{0})=\mathcal{S}(T_{h},h_{1})=\mathcal{S}_{1}$.
\item The process $b\rightarrow c$ is an isothermal process, with $\mathrm{d}T=0$.
The heat absorbed during this process is
\begin{alignat}{1}
Q_{bc}= & T_{h}\left[\mathcal{S}(T_{h},h_{1})-\mathcal{S}(T_{h},h'_{1})\right]\nonumber \\
= & T_{h}\left[\mathcal{S}(T_{h},h_{1})-\mathcal{S}(T_{l},h_{0})\right]\\
= & T_{h}\left(\mathcal{S}_{1}-\mathcal{S}_{0}\right)=Q_{in}>0,
\end{alignat}
here $\mathcal{S}(T_{l},h_{0})=\mathcal{S}_{0}$. 
\item The process $c\rightarrow d$ is adiabatic, so no heat is exchanged
$Q_{cd}=0$. The work done during this process becomes
\begin{alignat}{1}
W_{cd}= & -\Delta{\cal E}_{cd}=-({\cal E}_{d}-{\cal E}_{c})\nonumber \\
= & {\cal E}(T{}_{h},h_{1})-{\cal {\cal E}}(T_{l},h'_{0}).
\end{alignat}
\item The process $d\rightarrow a$ is isothermal, with ${\rm d}T=0$. The
heat released during this process is expressed by
\begin{alignat}{1}
Q_{da}= & T_{l}\left[\mathcal{S}(T_{l},h_{0})-\mathcal{S}(T_{l},h'_{0})\right]\\
= & T_{l}\left[\mathcal{S}(T_{l},h_{0})-\mathcal{S}(T_{h},h_{1})\right]\\
= & T_{l}\left(\mathcal{S}_{0}-\mathcal{S}_{1}\right)=Q_{out}<0.
\end{alignat}
Therefore, the work done during closed reversible cyclic becomes 
\begin{alignat*}{1}
W_{net}= & Q_{bc}+Q_{da}.
\end{alignat*}
The corresponding heat absorbed when the system\textquoteright s entropy
changes from $\mathcal{S}_{0}$ to $\mathcal{S}_{1}$ is given by
\begin{equation}
Q_{in}=Q_{bc}=T_{h}\left(\mathcal{S}_{1}-\mathcal{S}_{0}\right).
\end{equation}
\end{enumerate}
When the system operates as an engine, the corresponding heat engine
efficiency is described by
\begin{alignat}{1}
\eta_{c}= & \frac{W_{net}}{Q_{bc}}=1+\frac{Q_{da}}{Q_{bc}}=1-\frac{T_{l}}{T_{h}}=1-\frac{T_{p_{1}}}{T_{p_{2}}}.
\end{alignat}
In contrast, when the system operates in other engine modes, the corresponding
thermal efficiency is defined in Table \ref{tab:engines}. 

\subsection{Quantum many-body Otto machine cycle}

\begin{figure}
\includegraphics[scale=0.9]{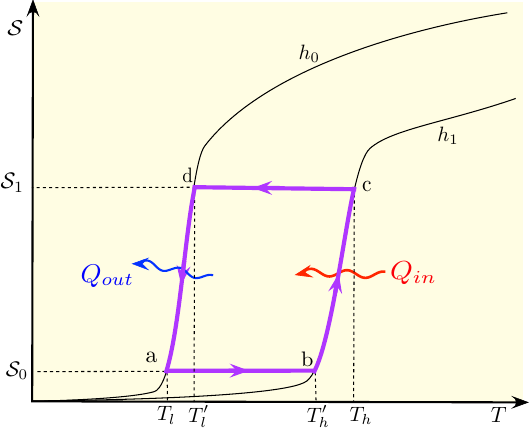}\caption{\label{fig:Otto-cycl} Schematic representation of quantum many-body
Otto machine, assuming $T_{l}<T'_{l}$ with $T_{l}<T_{p_{1}}$ and
$T'_{h}<T_{h}$ with $T'_{h}<T_{p_{2}}$.}
\end{figure}

Now, let us analyze a typical reversible Otto engine cycle. A schematic
representation of the process is depicted in Fig. \ref{fig:Otto-cycl}.

\begin{enumerate}
\item The process $a\rightarrow b$ is adiabatic, meaning no heat is transferred
$Q_{ab}=0$. Let us assume a constant entropy, $\mathcal{S}(T_{l},h_{0})=\mathcal{S}(T'_{h},h_{1})=\mathcal{S}_{0}$.
Using the first law of thermodynamics, the work done during this process
is given by
\begin{alignat}{1}
W_{ab}= & -\Delta{\cal E}_{ab}=-({\cal E}_{b}-{\cal E}_{a})\nonumber \\
= & \mathcal{E}(T_{l},h_{0})-\mathcal{E}(T'_{h},h_{1}),
\end{alignat}
\item The process $b\rightarrow c$ is an isomagnetic (constant magnetic
field) process, thus ${\rm d}h=0$. This implies the work is null.
Therefore, the heat transferred becomes 
\begin{alignat}{1}
Q_{bc}= & \Delta\mathcal{E}_{bc}=\mathcal{E}(T_{h},h_{1})-\mathcal{E}(T'_{h},h_{1})=Q_{in}>0.\label{eq:Q_bc-otto}
\end{alignat}
\item The process $c\rightarrow d$ follows an adiabatic process, which
means there is no heat transferred $Q_{cd}=0$. By fixing constant
entropy at $\mathcal{S}(T_{h},h_{1})=\mathcal{S}(T'_{l},h_{0})=\mathcal{S}_{1}$,
only work is performed during this process
\begin{alignat}{1}
W_{cd}= & -\Delta{\cal E}_{cd}=-({\cal E}_{d}-{\cal E}_{c})\nonumber \\
= & \mathcal{E}(T_{h},h_{1})-\mathcal{E}(T'_{l},h_{0}).
\end{alignat}
\item The process $d\rightarrow a$ is an isomagnetic process, meaning ${\rm d}h=0$.
This implies that no work is performed during this process. Therefore,
when energy changes occur, only heat is transferred
\begin{alignat}{1}
Q_{da}= & \Delta\mathcal{E}_{da}=\mathcal{E}(T_{l},h_{0})-\mathcal{E}(T'_{l},h_{0})=Q_{out}<0.\label{eq:Q_da-otto}
\end{alignat}
\end{enumerate}

\section{\label{sec:Otto-Machine-efficiency}Otto Machine efficiency}

In this Appendix, we analyze the efficiency of the Otto machine around
the pseudo-transition region. At the pseudo-critical temperature,
the specific heat exhibits a very sharp and narrow peak. Therefore,
the Taylor expansion around the pseudo-critical temperature $T_{p}$
can be written as follows

\begin{equation}
C=C_{p}+\frac{1}{2}C_{p}^{(2)}(T-T_{p})^{2}+\cdots
\end{equation}
Here, we have that $C_{p}^{(1)}=\left(\frac{\partial C}{\partial T}\right)_{T_{p}}=0$,
which corresponds to the maximum of the specific heat (not to be confused
with the specific heat at constant pressure), while $C_{p}^{(2)}=\left(\frac{\partial^{2}C}{\partial T^{2}}\right)_{T_{p}}<0$.
On the other hand, the following relation holds
\begin{equation}
\left(\frac{\partial\mathcal{S}}{\partial T}\right)=\frac{C}{T}.
\end{equation}
Integrating around $T\rightarrow T_{p}$, we can obtain the entropy
as follows 

\begin{alignat}{1}
\intop_{\mathcal{S}_{p}}^{\mathcal{S}}{\rm d}\mathcal{S}= & \intop_{T_{p}}^{T}\frac{1}{T}\left[C_{p}+\frac{1}{2}C_{p}^{(2)}(T-T_{p})^{2}\right]{\rm d}T\nonumber \\
\mathcal{S}-\mathcal{S}_{p}= & C_{p}\ln\left(\frac{T}{T_{p}}\right)+\frac{1}{4}C_{p}^{(2)}\left(T-T_{p}\right)\left(T-3T_{p}\right)\nonumber \\
 & +\frac{1}{2}C_{p}^{(2)}\ln\left(\frac{T}{T_{p}}\right),\label{eq:Sp-Tp-apprx}\\
\mathcal{S}-\mathcal{S}_{p}= & \left(C_{p}+\frac{T_{p}^{2}}{2}C_{p}^{(2)}\right)\ln\left(\frac{T}{T_{p}}\right)\nonumber \\
 & +\frac{1}{4}C_{p}^{(2)}\left(T-T_{p}\right)\left(T-3T_{p}\right).
\end{alignat}

If we wish to express the temperature analytically from equation \eqref{eq:Sp-Tp-apprx},
we encounter a transcendental equation that must be solved numerically.
However, for our purposes, we will consider only the first-order approximation
as the relevant term, from which the temperature can be expressed
analytically. Therefore, from \eqref{eq:Sp-Tp-apprx}, we obtain the
following simple expression

\begin{equation}
\mathcal{S}=\mathcal{S}_{p}+C_{p}\ln\left(\frac{T}{T_{p}}\right).
\end{equation}
Now, we can express the temperature as a function of entropy
\begin{alignat*}{1}
T= & T_{p}{\rm e}^{\left(\mathcal{S}-\mathcal{S}_{p}\right)/C_{p}}.
\end{alignat*}
From these relations, we can express the cold temperatures as follows
\begin{alignat*}{1}
T_{l}= & T_{p_{1}}{\rm e}^{\left(\mathcal{S}_{0}-\mathcal{S}_{p_{1}}\right)/C_{p_{1}}}\\
T'_{l}= & T_{p_{1}}{\rm e}^{\left(\mathcal{S}_{1}-\mathcal{S}_{p_{1}}\right)/C_{p_{1}}.}
\end{alignat*}
Substituting these into the heat released $Q_{da}$ during the d-a
process of the Otto cycle (as given in equation \eqref{eq:Q_da-otto}),
we obtain:
\begin{alignat}{1}
Q_{da}= & C_{p_{1}}T_{p_{1}}\left[{\rm e}^{\mathcal{S}_{0}/C_{p_{1}}}-{\rm e}^{\mathcal{S}_{1}/C_{p_{1}}}\right]{\rm e}^{-\mathcal{S}_{p_{1}}/C_{p_{1}}}\nonumber \\
\approx & T_{p_{1}}\left(\mathcal{S}_{0}-\mathcal{S}_{1}\right)\left[1+\tfrac{1}{2C_{p_{1}}}\left(\mathcal{S}_{0}+\mathcal{S}_{1}-2\mathcal{S}_{p_{1}}\right)\right].
\end{alignat}
Similarly, we can express the entropies for two fixed hot temperature
reservoirs as follows
\begin{alignat*}{1}
T'_{h}= & T_{p_{2}}{\rm e}^{\left(\mathcal{S}_{0}-\mathcal{S}_{p_{2}}\right)/C_{p_{2}}}\\
T_{h}= & T_{p_{2}}{\rm e}^{\left(\mathcal{S}_{1}-\mathcal{S}_{p_{2}}\right)/C_{p_{2}}}.
\end{alignat*}
Analogous to the previous result, the heat absorbed $Q_{bc}$ during
the b-c process (as described by equation \eqref{eq:Q_bc-otto}) becomes
\begin{alignat}{1}
Q_{bc}= & C_{p_{2}}T_{p_{2}}\left[{\rm e}^{\mathcal{S}_{1}/C_{p_{2}}}-{\rm e}^{\mathcal{S}_{0}/C_{p_{2}}}\right]{\rm e}^{-\mathcal{S}_{p_{2}}/C_{p_{2}}}\nonumber \\
\approx & T_{p_{2}}\left(\mathcal{S}_{1}-\mathcal{S}_{0}\right)\left[1+\tfrac{1}{2C_{p_{2}}}\left(\mathcal{S}_{0}+\mathcal{S}_{1}-2\mathcal{S}_{p_{2}}\right)\right].
\end{alignat}

Therefore, the ratio between heat released and absorbed is
\begin{alignat}{1}
\frac{Q_{da}}{Q_{bc}}\approx & -\frac{T_{p_{1}}}{T_{p_{2}}}\frac{\left(1+\tfrac{1}{2C_{p_{1}}}\left(\mathcal{S}_{0}+\mathcal{S}_{1}-2\mathcal{S}_{p_{1}}\right)\right)}{\left(1+\tfrac{1}{2C_{p_{2}}}\left(\mathcal{S}_{0}+\mathcal{S}_{1}-2\mathcal{S}_{p_{2}}\right)\right)},\nonumber \\
\approx & -\frac{T_{p_{1}}}{T_{p_{2}}}\left(1-\tfrac{1}{2C_{p_{2}}}\left(\mathcal{S}_{0}+\mathcal{S}_{1}-2\mathcal{S}_{p_{2}}\right)\right),\nonumber \\
\approx & -\frac{T_{p_{1}}}{T_{p_{2}}}+\frac{T_{p_{1}}}{T_{p_{2}}C_{p_{2}}}\left(\frac{\mathcal{S}_{0}+\mathcal{S}_{1}}{2}-\mathcal{S}_{p_{2}}\right),
\end{alignat}
since $C_{p_{1}}$ is considerable larger than $C_{p_{2}}$, the last
term is positive. From this result, one can derive the results given
in eq.\eqref{eq:Otto-effcy}.

\end{document}